\documentclass{llncs}
\usepackage{caption} 
\usepackage{subcaption}
\usepackage{graphicx}
\usepackage{wrapfig}
\usepackage{tikz}
\usepackage[T1]{fontenc}
\usepackage{url}
\usepackage{multicol}
\usepackage{footmisc}

\usepackage[sectionbib,numbers,sort&compress]{natbib}
\graphicspath{{./images/}}
\usepackage{parskip}
\usepackage{caption}
\usepackage[hyperfootnotes=false]{hyperref}
\begin{document}

\title{Towards MatCore: A Unified Metadata Standard for Materials Science}
%
%
\author{Jane Greenberg\inst{1}
\and P\'{a}mela B\'{o}veda-Aguirre\inst{2}
\and John Allison\inst{3}
\and Pietro Asinari\inst{4}
\and Maria Chan\inst{5}
\and Anand Chandrasekaran\inst{6}
\and Elif Ertekin\inst{7}
\and Emmanouel Garoufallou\inst{8}
\and Giulia Galli\inst{9}
\and Paolo Giannozzi\inst{10}
\and Feliciano Giustino\inst{11}
\and Gerhard Goldbeck\inst{12}
\and Hendrik Heinz\inst{13}
\and Arthi Jayaraman\inst{14}
\and Vincenzo Lordi\inst{15}
\and Kristin A. Persson\inst{16}
\and Gian-Marco Rignanese\inst{17}
\and Aidan Thompson\inst{18}
\and Eric Toberer\inst{19}
\and  Scott McClellan\inst{1}
\and Ellad B. Tadmor\inst{2}
}
\institute{Metadata Research Center, Drexel University,
\email{jg3243@drexel.edu}
\and University of Minnesota, \email{tadmor@umn.edu}
\and University of Michigan, Ann Arbor
\and Politecnico di Torino, Italian National Institute of Metrological Research 
\and Argonne National Laboratory
\and Schrödinger, Inc.
\and University of Illinois, Urbana-Champaign
\and MetaDATA LAB, International Hellenic University
\and University of Chicago
\and DMIF, Universit\`{a} di Udine, CNR - IOM
\and University of Texas, Austin
\and Goldbeck Consulting, EMMC
\and University of Colorado, Boulder 
\and University of Delaware
\and Lawrence Livermore National Laboratory
\and University of California, Berkeley 
\and UCLouvain, OPTIMADE
\and Sandia National Laboratory
\and Colorado School of Mines
}



%
\authorrunning{Greenberg et al.}
%
%
\maketitle              
\begin{abstract}
The materials science community seeks to support the FAIR principles for computational simulation research. The MatCore Project was recently launched to address this need, with the goal of developing an overall metadata framework and accompanying guidelines. This paper reports on the MatCore goals and overall progress. Historical background context is provided, including a review of the principles underlying successful core metadata standards. The paper also presents selected MatCore examples and discusses future plans.

\end{abstract}

\section{Introduction}
Materials science is an interdisciplinary field that draws from physics, engineering, biology, mathematics, and other intersecting disciplines \citep{bensaude2001construction, callister2020fundamentals, national1989materials}. Materials scientists study matter, specifically the relationship between the atomic or molecular level of a material's structure and its associated properties. The overall goal is to design and develop new materials, or improve the performance of existing materials. For example, a process for producing metal alloys may be modified to improve corrosion resistance.

Materials scientists use a range of research techniques spanning experimental, in-the-lab activities to modeling and simulation. These approaches along with technical and computational advances have radically increased the amount of materials science research data that is generated on a daily basis. This growth has further introduced unprecedented challenges and opportunities as researchers seek to effectively manage their research data, support the FAIR (Findable, Accessible, Interoperable, and Reusable) data principles \citep{wilkinson2016fair}, and apply machine learning (ML) and artificial intelligence (AI) techniques.

Metadata has taken on a an increasingly significant role in concert with these changes, as small-scale university research groups to large-scale industry laboratories mandate the use of metadata standards. As a result, materials scientists have adopted and modified existing metadata standards, and formed groups to develop new standards. Although these activities have advanced metadata practices and various research infrastructures, they lack a cohesive framework. Consequently, materials scientists seeking to work with  metadata standards are challenged on where to begin. The materials science community needs a scaffolding that provides a base-level entry point and which also contains core components to facilitate connectedness across the field. Development of a framework containing core standards supporting discovery and other metadata functions is particularly important for those engaged in computational simulation given unprecedented opportunities occurring in the field of AI \citep{pyzer2022accelerating}.

This need underlies the \emph{Materials Core Metadata} (MatCore) project, which was launched in early 2024. MatCore includes seven working groups organized across a unified framework. The aim is to develop a set of core metadata standards and implementation guidelines supporting key area of computational materials science research. This paper reports on the MatCore work. First, by way of background the paper describes historical metadata developments, identifies several important materials science metadata approaches, and reviews a number of core metadata standards. Next, the paper identifies MatCore’s goals and objectives and describes the standard structure. This is followed by the initial process of establishing the effort (Phase 0) and the development of the initial standard (Phase 1) including some examples. The last section summarizes MatCore progress and identifies next steps.
\section{Background Context}
\subsection{Materials Science Metadata}
Metadata, while primarily viewed as a digital asset, may exist as a physical artifact, or a digital record of a physical object. Consider the Greek astronomer Hipparchus of Nicaea, who compiled the first known stellar catalog in the second century BCE \citep{goldstein1991introduction}. He recorded positions of stars and their celestial coordinates, essentially capturing both data and metadata about stars observed in our stellar environs. Throughout history materials science researchers have shared their ``recipes'' for material synthesis, first through oral tradition and then the written word. For example, the Kaogong Ji, a technical encyclopedia, written in western China between the 3rd and 5th centuries BCE, contains recipes for bronze casting \citep{luo2020tentative}. Another example is the Leyden Papyrus found in Thebes, Egypt, and written in the 4th century CE \citep{hunt1976oldest}. This metallurgical handbook includes recipes for gilding silver, formulating base metal alloys, and soldering gold, some of which are drawn from older works.

The exact starting point for digital, structured materials science metadata is difficult to determine, although such activities began in the early 1960's with database development establishing key--value pairs in relational databases,  and the transition to hypertextual environments \citep{ghiringhelli2023shared, westbrook1991role}. Data interoperability presented a significant challenge \citep{stanton1991computerization} and underscored the need for solutions. The American Society for Testing Materials (ASTM) committee E49, ``Computerization of Material Property Data'' \citep{rumble1991standards} recommendations present another metadata milestone. The recommendations, released in 1985, advocated for researchers to include standard descriptors, identifiers, characterization data, and other features in materials databases. Roughly a decade later we see the adoption of standardized markup languages, such as the eXtensible Markup Language (XML), which underlies the materials markup languange (MatML) developed by NIST \citep{varde2006matml} and ontology design, supported by RDF/XML. Additionally, the Crystallographic Information Framework (CIF) format for crystallographic data was introduced in 1991 \citep{hall1991crystallographic}, and serves as an exemplary metadata standard adapting to disciplinary change. Indeed, the CIF format is a well-established standard, although views vary on if it is a metadata standard or a data standard.

Today, materials scientists can further draw from repositories, such as the Digital Curation Center's Disciplinary Metadata Directory (DCC/DMD), which provides access to an array of metadata standards. The DCC/DMD categorizes disciplinary metadata standards, profiles, tools, and use cases \citep{ball2016metadata} under the following five categories: \textit{general research},  \textit{physics},  \textit{biology},  \textit{earth science}, and \textit{social science \& humanities}, and many of the schemes registered are relevant to various areas of materials science. Materials science researchers also have access to ontologies and other semantic systems through resources, such as the Industrial Ontology Foundry \citep{oag_industrial_ontology_foundry}, MatPortal \citep{fraunhofer_matportal}, and NOMAD \citep{scheffler2022fair}. Collectively, these metadata developments have advanced materials science metadata practices. Despite this progress, the dispersed standards environment presents boundaries \citep{ghiringhelli2023shared} that impede materials scientists' full embrace and use of metadata. A look at core standards within longer-standing cohesive metadata environments can provide guidance for addressing this challenge and better situating MatCore activities.

\subsection{Core Metadata Standards: Motivation and Success}
The course of metadata history has included a series of core standards developed by disciplinary community members. The most successful of these efforts embody the spirit of \textit{open science, transparency}, and \textit{community ownership}. These factors are grounded in the \textit{open-source movement}, which many metadata developers also traversed.

One of the earliest core metadata standards is the Internet Anonymous FTP Archives (IAFA), developed by the Internet Engineering Task Force (IETF) and released in 1995 \citep{beckett1995iafa, deutsch1995publishing}. IAFA offered a suite of metadata templates for publishing and exchanging of anonymous FTP information via Gopher, such as text indices, Linux Software Maps (LSMs), and other objects. Members of the IAFA community were instrumental in developing the Dublin Core, which is arguably one of the best known, interdisciplinary core metadata standards  \citep{weibel2000dublin, weibel1997element, arakaki2018dublin}. Initiated in March 1995 at a workshop co-hosted by the National Center for Super Computing Applications (NCSA), the Dublin Core principles aim to support metadata \textit{simplicity},  \textit{interoperability},  \textit{modularity}, and \textit{extensibility}. Dublin Core has had a global impact, and many of today's frequently used metadata standards map their core properties to the this standard.

Additional core standards include the VRA Core \citep{loc_vra_core} for describing images of art and artifacts, the Darwin Core \citep{wieczorek2012darwin} for scientific, primarily biological specimens and samples, and the minimum information standards spearheaded by the Minimum Information for Biological and Biomedical Investigations (MIBBI) project \citep{taylor2008promoting}. Darwin Core has a series of extensions, such as the Audiovisual Core (formerly Audubon Core) \citep{morris2013discovery} for describing multimedia collections related to biodiversity. The MIBBI guidelines also includes a suite of standards (e.g., Minimum Information About a Plant Phenotyping Experiment (MIAPPE) \citep{pommier2024long}, Minimum Information About a Microarray Experiment (MIAME)\citep{brazma2001minimum}, etc.) to guide reporting. All core standards noted here have been developed in open, community-driven environments and support the FAIR principles, even prior to their publication. Indeed, reviewing the full spectrum of core metadata standards supporting community connections is beyond the scope of this paper. What is important is to recognize that the success of core metadata standards hinges on open, community-driven approaches. The examples here have presented RFC (request for comment) documents, a requirement for formal standard endorsement. Additionally, many of these standards adhere to the ISO 11179, Metadata registries (MDR) standard \citep{iso_iec_11179_1}. Overall, lessons learned from these examples both inform and motivate the work being pursued with the MatCore project.

\section{MatCore: Goals and Objectives}
\label{sec:objectives}

MatCore's overriding goal is to support the FAIR principles for computational materials science research. In order for the data generated by computer simulations of materials to be useful, they must be accompanied by information that fully characterizes the nature of the computation performed and the material being modeled. Further, to enable researchers to reproduce generated results, the specific parameters and settings input to the simulation program must be provided.
Interoperability is also critical, given the goal to publish simulation data and information in open repositories for the purpose of collaboration. An additional, unifying MatCore goal is to build and sustain a metadata framework that facilitates a more cohesive, community-driven metadata approach.

Specific objectives being addressed in the Phase~1 working groups (WGs) include the development of core metadata standards covering the following five computational materials science methods:
\begin{enumerate}
\item Density Functional theory (DFT): First-principles computational methods based on quantum mechanics for predicting the ground state structure and properties of materials.
\item Classical Molecular Dynamics (MD): Methods for integrating the equations of motion of atoms using approximate fitted models for atomic interactions to predict classical static and dynamics properties of materials.
\item Many-Body Perturbation Theory (MBPT): First-principles computational methods based on quantum mechanics (such as $GW$ and BSE) for computing properties of materials involving excited states.
\item Machine Learning (ML): Data-driven approaches that employ ML techniques to predict material structure and properties.
\item Derivative Methods: Hybrid calculations involving mixtures of other computational methods to predict material properties.
\end{enumerate}

Additionally, there is a Mimimal Metadata WG that is working on defining a core set of common metadata properties required for every computational method.  Finally, there is also a Metadata Implementation WG focused on identifying best practice recommendations to assure metadata quality.

\section{MatCore Standard}
\begin{figure}[t]
\resizebox{\textwidth}{!}{%
\begin{tikzpicture}
\tikzset{
    rectangle/.style = {draw, text width=2.3cm, minimum height=0.8cm},
    style1/.style = {rectangle, rounded corners=2pt, thin, align=center, fill=pink!30},
    style2/.style = {rectangle, rounded corners=2pt, thin, align=center, fill=pink!50},
    style3/.style = {rectangle, thin, align=left, fill=pink!50}
}
\tikzstyle{level 1} = [sibling distance=30mm] 
\tikzstyle{edge from parent} = [->,draw, >=latex]

\node[style1] {\small   Minimal MatCore Metadata}
    child {node[style2] (c1) {\small DFT\\ Method-Specific Metadata}}
    child {node[style2] (c2) {\small MD\\ Method-Specific Metadata}}
    child {node[style2] (c3) {\small GW/BSE\\ Method-Specific Metadata}}
    child {node[style2] (c5) {\small ML\\ Method-Specific Metadata}}
    child {node[style2] (c4) {\small Derivative\\ Method-Specific Metadata}};

\end{tikzpicture}

} 
\caption{Schematic of the MatCore structure.}
\label{fig:matcore:structure}
\end{figure}
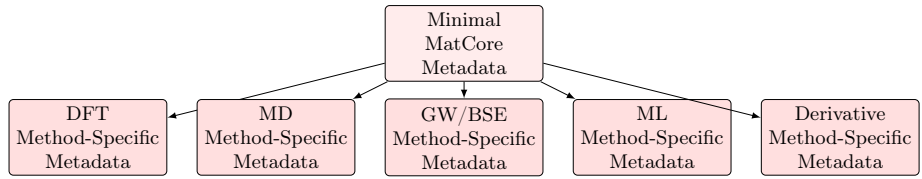

MatCore defines required and optional metadata to accompany datasets generated through computational materials science techniques that will allow researchers to understand, use, and, if desired, reproduce the data. The structure of the MatCore standard is based on a two-tier hierarchy (see Fig.~\ref{fig:matcore:structure}). The top level comprises the \textit{Minimal MatCore Metadata}, and the second level consists of templates of specialized metadata for key computational materials science methods.
\begin{enumerate}
\item Minimal MatCore Metadata \\
 The core component of the standard is the ``Minimal Matcore Metadata.'' Capturing this information is required for all datasets to specify the basic characteristics of the material being modeled and the performed computation. These general metadata properties apply to all computational methods.

\item Method-Specific Metadata \\
In addition to the core component, each dataset may optionally be accompanied by method-specific metadata. This secondary level provides detailed information in the form of method-specific parameters and settings to allow experts to better understand the nature of the computation.
\end{enumerate}

While the above presents the current structure for the MatCore standard, it may evolve over time. In particular, it is anticipated that additional method-specific WGs will form for existing and new computational method.

\section{The MatCore Standard Development Process}
Establishment of a metadata standard for computational materials science requires careful planning and building community support. This will be achieved through a series of phases:
\begin{itemize}
\item Phase 0: Establishment of the MatCore Standards Committee and Funding Recruitment
\item Phase 1: Development of the Draft MatCore Standard
\item Phase 2: Community Request for Comment
\item Phase 3: MatCore Committee Hearings and definition of the Final MatCore Standard
\item Phase 4: Reporting and Planning for Ongoing MatCore Standard Support
\end{itemize}
At the time of writing, Phase~0 has been completed, and Phase~1 is underway, as described below.

\subsection{Phase 0: Establishment of the MatCore Standards Committee and Funding Recruitment}
Phase~0 of the MatCore standard began in January 2023.
The development of MatCore standard requires the collective expertise of leading researchers in the computational materials science methods discussed in Section~\ref{sec:objectives}. To this end, over a period of about a year, a committee comprised of the authors of this paper was established and engaged in discussions to establish a preliminary standard design.

A key consideration in establishment in the MatCore Committee was to assemble a group of researchers with expertise that spans that topics to be covered by the standard as well as in metadata standard development. Although, the focus of the current effort is on computational methods for solid-state materials, it was considered important to include both experimentalists as well as researchers working on soft matter on the Committee to benefit from their experience. With this broad input, it is hoped that a more robust standard will be developed that can be expanded to support computational materials science for both hard and soft matter, and ultimately experimental materials science, which is a far more difficult problem.

Following the established of the MatCore Committee and the completion of a preliminary standard design, a proposal was submitted to the U.S.\ National Science Foundation (NSF) by authors Tadmor, Persson and Giustino to support the development of the MatCore Standard. NSF funding was received from the Division of Materials Research (DMR), and work on Phase~1 has been initiated.

\subsection{Phase 1: Development of the Draft MatCore Standard}
Phase~1 of MatCore Standard development began in June 2024 and involves a series of steps that are currently underway. Fig.~\ref{fig:process} presents the development process in Phase~1. Step 1, involves a historical review that is being performed in parallel to steps 2, 3, and 4 that are focused on the definition of the metadata structure:

\begin{figure}[b]
\includegraphics[scale=0.85]{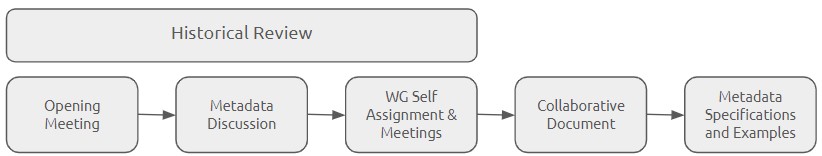}
\caption{Phase~1 of the MatCore standard development Process.}
\label{fig:process}
\end{figure}


\begin{enumerate}
\item \textit{Historical Review:}
A historical review is being conducted to identify and study previous efforts, frameworks, standards, and projects aimed at providing FAIR access to data in materials science.
\item \textit{Kickoff:}
A series of kickoff meetings were held for the MatCore WGs to provide an overview to new MatCore Commitee members and discuss next steps.
\item \textit{Metadata Primer:}
A talk by an expert on metadata theory, followed by an open discussion, was held for the MatCore Committee. The talk presented metadata definitions, outlined the standards development process, reviewed the FAIR principles, and shared recommendations for the MatCore project.
\item \textit{Working Group Self-Assignment and Meetings:}
MatCore Committee Memebers selected which general and method-specific WGs to participate in (see Section~\ref{sec:objectives}) based on their expertise, experience and preference. Regular meetings were arranged for each WG. The minimal metadata and method-specific WGs discussed which metadata needed to be included, and the implementation WG focused on metadata best practices for the MatCore standard.
\item \textit{Collaborative Development of MatCore Metadata Standard:}
A collaborative document was shared with all participants, which allows members to view and contribute to each others activities. As a result, each WG benefited from both its own members' expertise, as well as that of the other WGs.
\end{enumerate}

\begin{figure}[t]
\includegraphics[scale=0.70]{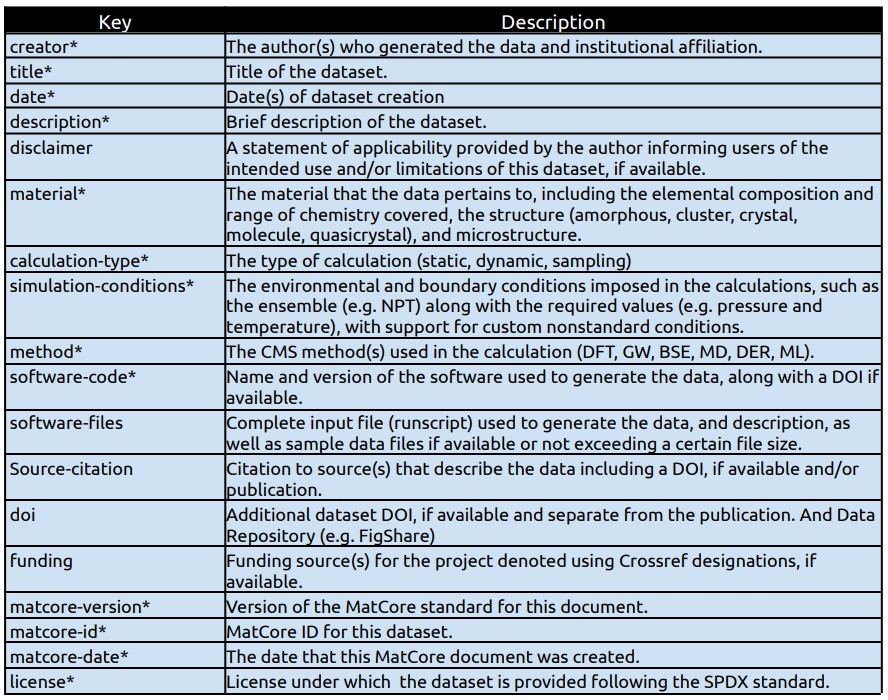}
\caption{Preliminary draft of the Minimal MatCore Metadata specification. Starred keys are required.}
\label{fig:matcore:minimal}
\end{figure}

Over the last few months since the beginning of Phase~0, the MatCore WG's have been holding virtual meeting on an ad hoc to collaborate on their assignments. These meetings are difficult to arrange due to the range of time zones with participants across the United States and in Europe, and are also challenging due to the range of backgrounds of the participants (engineers, physicists, chemists, computer scientists, data science, metadata, etc.).

\begin{figure}[t]
\includegraphics[scale=0.70]{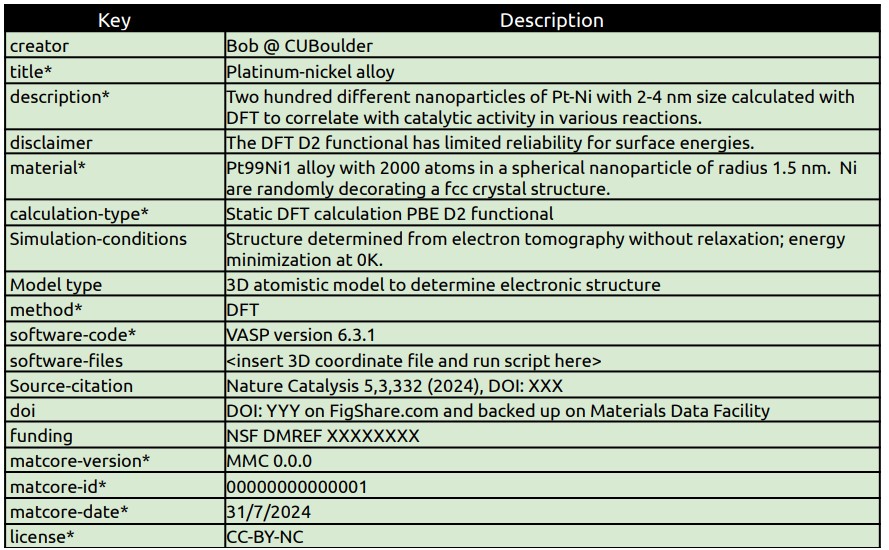}
\caption{Preliminary example of a Minimal MatCore Metadata specification for a dataset generated through DFT computations for a platinum-nickel alloy.}
\label{fig:matcore:minimal:example}
\end{figure}

\begin{figure}
\includegraphics[scale=0.70]{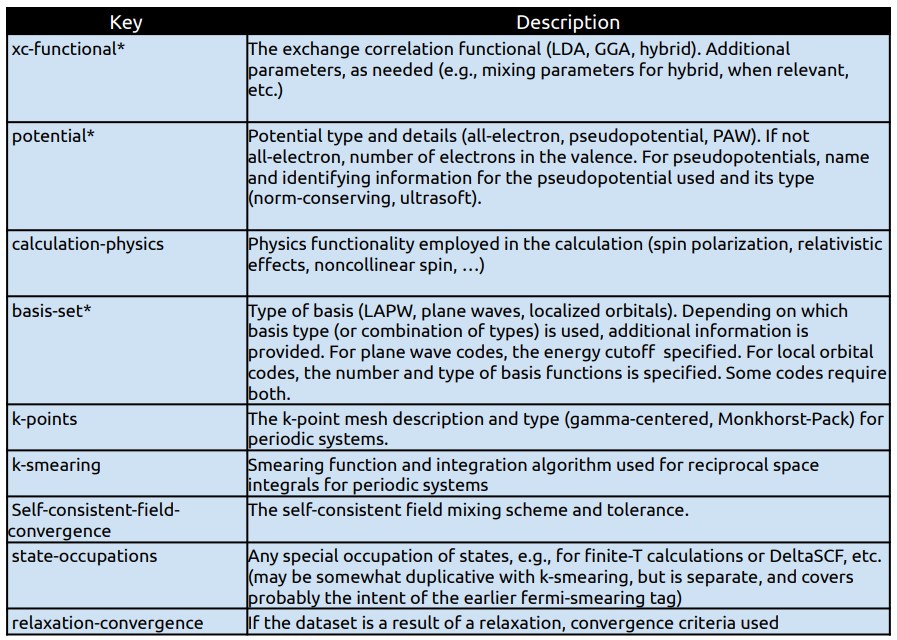}
\caption{Preliminary draft of the Method-Specific MatCore Metadata specification for DFT. Starred keys are required.}
\label{fig:matcore:dft}
\end{figure}

The Draft MatCore Standard, which is the end goal of Phase~1, is still under development, however significant progress has been made to date. Each WG has produced an initial draft of their metadata specification and the Implementation WG has identified several best practice recommendations. Additionally, several of the WG's have documented examples for their computational methods. Some of these are presented below. Fig.~\ref{fig:matcore:minimal} presents the Minimal MatCore Metadata specification, and Fig.~\ref{fig:matcore:minimal:example} provides an example of its application to a dataset generated by a DFT computation. Finally, Fig.~\ref{fig:matcore:dft} presents the DFT Method-Specific Metadata specification. Note that all of these specifications are preliminary and subject to change.

\section{Conclusion}

MatCore has been able to progress due to an engaged, open community of experts. The open approach and the shared MatCore development document has allowed for collaboration within and across all of the WGs. This is important as the materials researchers participating in MatCore may work more frequently with some methods, but they are knowledgeable across all methods. Additionally, MatCore includes those with expertise in standards development. This collaborative approach helps MatCore support the FAIR principles and ensures that computational materials science outputs are:
\begin{itemize}
    \item Findable, by providing clear identifiers and making them easily located.
    \item Accessible, by providing details about the methods, underlying code, and tools use.
    \item Interoperable, by using standardized formats, definitions and units.
    \item Reusable, by offering details for accurate replication of the resource.

\end{itemize}

Another important aspect of the MatCore project is the open, transparent, community-driven approach. This will allow for continued development and improvement over time. The work being conducted in Phase~1 will form a draft standard, which will be published online, and open for ublic comment in Phase~2. Feedback received will be evaluated at a MatCore Committee hearing in which selected community members will be invited to appear, and integrated into a revised version. The long-term goal is to establish the MatCore standard so that the materials science community can continue contributing to MatCore development and sustainability.

\section*{Acknowledgements}
The authors acknowledge partial support by the National Science Foundation (NSF) under grants DMR-2404283, OAC-2118201, and OAC-2320600. GG acknowledges funding from the European Union's Horizon 2020 research and innovation programme under grant agreement No.~953167 (OpenModel). PG acknowledges support from the European Union through the MaX Centre of Excellence for Supercomputing applications (project No. 101093374).


\bibliographystyle{splncs04}

\end{document}